# Performance of Quantum Key Distribution Protocol with Dual-Rail Displaced Photon States

Sergey A. Podoshvedov

*School of Computational Sciences, Korea Institute for Advanced Study, Seoul,130*-722

*e-mail:sap@kias.re.kr*

## ABSTRACT

We propose a scheme for quantum key distribution (QKD) protocol with dual-rail displaced photon states. Displaced single photon states carry bit value of code which may be extracted while coherent states carry nothing and they only provide inconclusive outcome. Developed QKD protocol works with experimental attendant noise to observe presence of malicious Eve. Pulses with large amplitudes unlike conventional QKD relying on faint laser pulses are used that may approximate it to standard telecommunication communication and may show resistance to eavesdropping even in settings with high attenuation. Information leakage to the eavesdropper is determined from comparison of output distribution of the outcomes with ideal one that is defined by two additional inaccessible to nobody, saving for who sends the pulses, parameters. Robustness to some possible eavesdropping attacks is shown.

## 1. Introduction

The quantum key distribution protocol provides a way for two remote parties (traditionally known as Alice and Bob) to share a secure random key by communicating over an open channel [1-5]. The two users have two kinds of communication channels at their disposal. One is a classical public channel that may be eavesdropped by any unauthorized person but cannot be modified and the second is a quantum channel. The quantum channel is used to transmit the secret key while the classical public channel is used to check possible presence of eavesdropping and to send the encoded message. Quantum mechanics ensures that any activities of potential eavesdroppers can be detected. If Alice and Bob are sure in security of their key, they finally process the obtained key (the raw key) to produce a much safer key (the final key) using classical methods of error correction and privacy amplification [6,7].

At present, there is a large collection of variations of QKD protocols [8]. Let us mention a few, chosen somewhat arbitrarily. The most famous QKD protocol is the four state scheme, usually referred to as the Bennet-Brassard 1984 (BB84) protocol. In this protocol, the transmission of a single photon randomly polarized along four directions is used [2]. The key idea of the BB84 protocol is that simultaneous measurements of noncommuting observables for a single photon in two conjugate bases are forbidden by quantum mechanics. In order words, the measurement of one observable made on the eigenstate of another observable inevitably introduces disturbance to the state. Eve has no any knowledge about the state sent by Alice and therefore she is forced on average half the time to introduce a disturbance into the state that can be detected as a bit error. One of possible variation of BB84 consists in using quantum systems of dimension greater than 2 [9]. Most of the existing schemes use an imperfect single-photon source since a single photon resource is difficult to produce experimentally, usually weak pulses were used in practice [10]. Such an implementation, in general case, may subject to the photon number-splitting attack [11]. To deal with imperfect source of single photons, many interesting methods was proposed [12] involving the decoy-state method [13].

Another possible way to implement secret sharing coding is based on use of pairs of Einstein-Podolsky-Rosen (EPR) correlated photons [3]. A communication protocol based on entangled pairs of qubits is presented in [14]. A system, which is conceptually the simplest, is the use of nonorthogonal quantum states [5]. Indeed, two nonorthogonal states cannot be distinguished unambiguously without perturbation only at the cost of some losses [15]. Initially, the two state protocol [5] was proposed to implement using interference of two classical pulses that is fragile under influence of decoherence.

Instead of use of single photons or weak coherent pulses, it is interesting idea any nonclassical field states are useful for quantum information processing and communication that was demonstrated on example of QKD with squeezed light [16]. Here, we propose to make use of nonclassical properties of the displaced single photon states to share secret coding between two sides. Displacement operator imposes additional varied degree of freedom on a photon state. According to studied model of QKD the inputs are the dual-rail displaced states rather than single photon $\left|1\right\rangle$ unlike [2]. In order words, carriers in the model are the optical pulses with different large amplitudes as in usual classical communication. The developed protocol of QKD is free of problems connected with interference. Let us also mention the displaced single photon state was experimentally generated in [17]. A possibility to conditionally generate displaced entangled states via nonlinear interaction of powerful pump beam with a crystal with $\chi^{(2)}$ nonlinearity was proposed in [18]. Another interesting application of the displaced states is the protocol of dense coding [19].

**2. Implementation of QKD with dual-rail displaced states**

In developed protocol, Alice prepares two ensembles of displaced states with different amplitudes of displacement (in general case, we deal with four states two of which are nonclassical). Every of the ensembles consists of only basis states, namely either product of displaced single photon and vacuum states or product of two displaced vacuum states (a two qubit system has four computational basis states denoted $|00\rangle$, $|10\rangle$, $|01\rangle$, and $|11\rangle$) but not their superpositions in the two-dimensional space. Basis elements $|00\rangle$, $|10\rangle$ with different amplitudes of displacement are not orthogonal to each other. Performance of the protocol is based on use of nonclassical properties of the displaced single photon states. On the receiving side, Bob has to distinguish between two displaced single photons $|10\rangle$ with different amplitudes for each incoming carrier. Since displaced single photon states are non-orthogonal, Bob cannot do it with certainty and he sometimes fails to extract correct outcome but once it gives one (this means that he performs a test by means of a generalized measurement known POVM [20]). Proposed protocol corresponds a communication channel known as a binary erasure channel with possible outcomes $0$, $1$, and $?$ ($?$ means inconclusive result) as in B-92 protocol [5]. For example, if Alice sends a $0$, Bob may get either a $0$ or an inconclusive result, but never $1$, saving for a case of Eve eavesdropping the communication channel sends herself $1$ by mistake. Inconclusive outcomes are also provided by sending of $|00\rangle$ or the same coherent states with different amplitudes.

Now, let us present mathematical details of the protocol. The states $|0,\alpha\rangle = \hat{D}(\alpha)|0\rangle$ and $|1,\alpha\rangle = \hat{D}(\alpha)|1\rangle$ are the displaced vacuum and one-photon states, respectively, where $\hat{D}(\alpha)$ is the displacement operator [18, 19]. A quantum system prepared by Alice is given by density matrix

$$\rho = \frac{1}{2}\rho_1 + \frac{1}{2}\rho_2, \tag{1a}$$

where

$$\rho_1 = P_1|\varphi_1\rangle\langle\varphi_1| + P_1^{'}|\varphi_1^{'}\rangle\langle\varphi_1^{'}|, \tag{1b}$$

$$\rho_2 = P_2|\varphi_2\rangle\langle\varphi_2| + P_2^{'}|\varphi_2^{'}\rangle\langle\varphi_2^{'}|, \tag{1c}$$

($P_1 + P_1^{'} = 1$ and $P_2 + P_2^{'} = 1$) with dual-rail displaced states defined as

$$|\varphi_1\rangle_{12} = |1,\alpha\rangle_1 |0,i\alpha\rangle_2, \tag{2a}$$

$$|\varphi_1^{'}\rangle_{12} = |0,\alpha\rangle_1 |0,i\alpha\rangle_2, \tag{2b}$$

$$|\varphi_2\rangle_{12} = |1,i\alpha_1\rangle_1 |0,\alpha_1\rangle_2, \tag{2c}$$

$$|\varphi_2^{'}\rangle_{12} = |0,i\alpha\rangle_1 |0,\alpha\rangle_2, \tag{2d}$$

where in general case $\alpha \neq \alpha_1$. The parameters $\alpha$, $\alpha_1$ and $P_1, P_1^{'}, P_2$, and $P_2^{'}$, respectively, are Alice's secret ones and they are hidden from both Bob and Eve. The states $|\varphi_1\rangle_{12}$ and $|\varphi_2\rangle_{12}$ may carry bit values of the coding ($0$ or $1$, respectively). We are going to call the states as bit ones. Since the states $|\varphi_1^{'}\rangle_{12}$ and $|\varphi_2^{'}\rangle_{12}$ do not carry any information to Bob, we call them disguised ones.

Bob prepares its measurement system as it is shown in Fig. 1 to extract some useful information from the states obtained from Alice. The measurement system involves a balanced beam splitter $\hat{B}_1$ described by the following matrix

$$B_1 = \frac{1}{\sqrt{2}}\begin{bmatrix} 1 & i \\ i & 1 \end{bmatrix}. \tag{3}$$

The outcome of the beam splitter (3) is the following

$$\hat{B}_1|\varphi_1\rangle_{12} = \frac{1}{\sqrt{2}}\left(|1\rangle_1 |0,i\sqrt{2}\alpha\rangle_2 + i|0\rangle_1 |1,i\sqrt{2}\alpha\rangle_2\right), \tag{4a}$$

$$\hat{B}_1|\varphi_1'\rangle_{12} = |0\rangle_1|0, i\sqrt{2}\alpha\rangle_2, \quad (4b)$$

$$\hat{B}_1|\varphi_2\rangle_{12} = \frac{1}{\sqrt{2}}\left(|1, i\sqrt{2}\alpha\rangle_1|0\rangle_2 + i|0, i\sqrt{2}\alpha\rangle_1|1\rangle_2\right), \quad (4c)$$

$$\hat{B}_1|\varphi_1'\rangle_{12} = |0, i\sqrt{2}\alpha\rangle_1|0\rangle_2. \quad (4d)$$

To achieve a discrimination of the outcomes (4a)-(4d) with off-the-shelf photon counters that can only differentiate between zero and more photons ($1,2,\ldots,n$), the simplest approach for Bob is to split them by two beam splitters as shown in Fig. 1. Consider a partial case of such a discrimination of a single photon and coherent state as it is shown in Fig. 2 (a, b). The beam splitter transforms a coherent state $|0, i\sqrt{2}\alpha\rangle$ to the product of two coherent states $|0, i\sqrt{2}\alpha\rangle_1|0\rangle_2 \rightarrow |0, i\alpha\rangle_1|0, i\alpha\rangle_2$, while a single photon to a superposition state $|10\rangle_{12} \rightarrow (1/\sqrt{2})(|10\rangle_{12} + |01\rangle_{12})$. If both detectors $D_1$ and $D_2$ register any photons, we know we detected a state $|0, i\sqrt{2}\alpha\rangle$. On the contrary, if either $D_1$ or $D_2$ does not click, we can assume with almost unity probability that was a single photon, especially in the case of large value of $\alpha$. The same is applicable to discrimination of all states in Eqs. (4a)-(4d). It follows from Eqs. (4a)-(4d) that three simultaneous clicks by detectors $D_1 - D_4$ in Fig. 1 are unambiguously identified as bit values ($0$ and $1$, respectively). All other events with three clicks less or more are identified as inconclusive outcome.

Given QKD protocol works as follows. Alice injects light in one of the four states (4a)-(4d) into a communication channel in random sequence. Via the use of a proposed detection system (Fig. 1) triggered on some photon statistics, presence of three simultaneous clicks in output Bob's statistics heralds the extraction of bit information. All the carries sent by Alice are numbered. Bob measures the incoming pulses to establish a one-to-one correspondence between sent and received pulses. At the point, where Bob may successfully extract bit value (three simultaneous clicks), they get perfectly correlated results. Bob has only to declare a number of the corresponding pulse (but not its result). All the rest inconclusive outcomes are discarded by Bob. This allows for Alice and Bob to share mutual information

$$I(A,B)=\log_2(P_1p_1+P_2p_2)-(P_1p_1\log_2(P_1p_1)+P_2p_2\log_2(P_2p_2))/(P_1p_1+P_2p_2), \quad (5)$$

where $p_1=p_2=0.5(1-P_0(\alpha))$ are the conditional probabilities for Bob to obtain a bit result provided that Alice sent $|\varphi_1\rangle_{12}$ and $|\varphi_2\rangle_{12}$, respectively, and $P_0(\alpha)=\exp(-2|\alpha|^2)+2\exp(-|\alpha|^2)(1-\exp(-|\alpha|^2))$, where $P_0(\alpha)$ is the probability to register three clicks less provided that the states were $|1\rangle_1|0,i\sqrt{2}\alpha\rangle_2$, $|0,i\sqrt{2}\alpha\rangle_1|1\rangle_2$. To simplify calculations in future consideration, we suppose $\alpha=\alpha_1$. Although, one should note the QKD protocol admits a possibility $\alpha\neq\alpha_1$ and even more Alice may vary amplitudes of every sent carriers provided that the phase relations of dual states remain constant. Given possibility may prevent the protocol from Eve's eavesdropping in the case of possible more skilful attacks that beyond our consideration. It is natural to assume that Alice delivers states $|\varphi_1\rangle_{12}$ and $|\varphi_2\rangle_{12}$ with equal probabilities $P_1=P_2=P$ that allows for Alice and Bob to share 1 bit of mutual information.

As well known, quantum cryptography cannot prevent eavesdropping, but any eavesdropping attempt can be detected by the legitimate users of the communication channel. This is related with that fact that eavesdropping affects the quantum state of the information carriers and results in an abnormal error rate. Therefore, before Bob publicly declares the number (but not the result of his measurement) where he successfully extracted a bit value, Alice and Bob have to test their communication channel by sacrificing a part of their data sufficient to estimate output distributions. Actually, there are three parameters to judge about possible presence of eavesdropping in the subset. The main such a parameter is the output distribution of bit and inconclusive outcomes given in the case of absence of eavesdropping by

$$P_0^{(Out)} = \frac{P_1}{4}\left(1 - P_0(\alpha)\right) = \frac{P}{4}\left(1 - P_0(\alpha)\right), \tag{6a}$$

$$P_1^{(Out)} = \frac{P_2}{4}\left(1 - P_0(\alpha_1)\right) = \frac{P}{4}\left(1 - P_0(\alpha)\right), \tag{6b}$$

$$P_?^{(Out)} = 1 - P_0^{(Out)} - P_1^{(Out)}. \tag{6c}$$

Note that neither Bob nor malicious Eve cannot know the output distribution of the bit and inconclusive outcomes since the parameters $P_1 = P_2 = P$ and $\alpha$ are chosen by Alice according to her own strategy and they are hidden from other participants. Eve can only listen to the talk between Alice and Bob through a public channel but she cannot correct the output distribution shared by Alice and Bob. Another important parameter whose change testifies presence of Eve in the communication channel is that which we call disguised probability $P_d$ being frequency of appearance of bit outcome while Alice sent one of the disguised states. The disguised states cannot give bit outcome only inconclusive outcome. The disguised probability $P_d$ has to be equal exactly zero in ideal case of absence of eavesdropping. Finally, Alice and Bob also may compare bit values of the chosen subset. For example, it is evident that single photon is not detected in mode $2$ if Alice sends a state $|\varphi_1\rangle_{12}$ and vise versa. Thus, these parameters may serve as indicators of presence or absence of the eavesdropping in the communication channel. If the parameters do not coincide with ideal then eavesdropping is detected and transmission is aborted. One should note it is possible directly to check a communication channel not sacrificing any subset of data. Indeed, Bob call corresponding number of his bit outcome and is it sufficient for Alice to estimate output distribution and disguised probability to compare it with Bob's. After that they can decide to take them or discard.

Displacing a state of light can be experimentally implemented by overlapping it with a strong coherent state $|0,\alpha_2\rangle$ upon a highly reflecting beam splitter. Now, we follow a method of [17] used to experimentally generate a displaced single photon. We suppose that the initial single photon is prepared by means of conditional measurements on a biphoton generated via parametric down conversion. It was discussed in [17] that imperfections associated with experimental technique result in the photon being prepared with a substantial admixture of the vacuum state $\rho_A = \eta|1\rangle\langle 1| + (1-\eta)|0\rangle\langle 0|$, where $\eta$ is the preparation efficiency. The preparation efficiency may accounts for the spontaneous parametric converter dark count events. In such an event, the quantum state in the output mode in not conditioned on that in used converter channel. Alice only needs to estimate the value of her preparation efficiency. She uses a beam splitter with arbitrary parameters $T$ and $R$ but which knows only she ($T$ and $R$ are transmittance and reflectance, respectively)

$B' = \begin{bmatrix} T & R \\ -R^* & T^* \end{bmatrix}$ to overlap her prepared state $\rho_A$ with a strong coherent field $|0,\alpha_2\rangle$. Indeed, Alice uses two coherent fields $|0,\alpha_2\rangle_{A_1}|0,\alpha_2'\rangle_2$ to provide outcome (1a). The output state in the beam splitter $B'$ is calculated by applying beam splitter transformation rules. A state in the modes 1 and 2 is obtained if we take trace upon states in auxiliary mode $A_1$. The beam splitter acts upon the incident single photon state simply as a lossy reflector, reducing its efficiency by a factor $|R|^2$. Also, the beam splitter causes the displacement of the state $\rho_A$, producing a final statistical mixture of displaced Fock states as

$$\begin{aligned}\rho_A' = \eta|R|^2\left(|1,\alpha_1 T\rangle_{11}\langle 1,\alpha_1 T|\right)\left(|0,\alpha_1'\rangle_{22}\langle 1,\alpha_1'|\right)+ \\ \left(\eta|T|^2+1-\eta\right)\left(|0,\alpha_1 T\rangle_{11}\langle 0,\alpha_1 T|\right)\left(|0,\alpha_1'\rangle_{22}\langle 0,\alpha_1'|\right)\end{aligned}$$

. The state $\rho_A'$ is the state $\rho_1$ (Eq. (1b)) provided that $P_1 = \eta|R|^2$, $P_2 = \eta|T|^2 + 1 - \eta$, $\alpha = \alpha_1 T$ and $i\alpha = \alpha_1'$. The same is applicable to generate $\rho_2$ (Eq. (1c)). Thus, unavoidable noise being in practice due to technical imperfections is substantial part of the developed protocol of key distribution unlike most of other protocols which require ideal resource of quantum states, for example, ideal resource of single photons. Any of unauthorized observers may estimate preparation efficiency $\eta$ but it is hardly possible for him to guess the reducing factor $|R|^2$ and all the more the values of parameters $\alpha$ and $\alpha_1$ that initially are known only to Alice. The additional modulation of the outcome of conditioned down converter gives a possibility Alice to use quantum states with additional secret parameters to transmit them to Bob.

**3. Robustness to eavesdropping**

We now analyze some of eavesdropping strategies. Note that direct measurement of incoming pulse does not give answer which of the four states was sent. If Eve prefers to measure dependence of falling field on the relative phase she may use a scheme that involves homodyning the signal field with a reference signal known as the local oscillator before photodetection. Homodyning with a reference signal of fixed phase gives the phase sensitivity necessary to yield the quadrature variances. A measurement of quadrature components shows that statistical characteristics $\langle 0,\alpha|\hat{a}|0,\alpha\rangle = \langle 1,\alpha|\hat{a}|1,\alpha\rangle = \alpha$ are equal. Then, Eve may not be aware of which type of state she has (bit or disguised) if she measured a definite value of the quadrature component ($\langle 0,\alpha|\hat{X}|0,\alpha\rangle = \langle 1,\alpha|\hat{X}|1,\alpha\rangle$).

The most practical eavesdropping strategy may be intercept-resend attack. Eve intercepts the quantum carrier on its way from Alice and Bob and performs the same measurement as it does Bob, namely, using a beam splitter $B_1$ (Eq. (3)). After the measurement, Eve sends to Bob another quantum carrier in one of the four states (2a)-(2d), looking at her outcome and following some chosen strategy. Eve's strategy may be the following. If Eve obtains bit value then she again sends the corresponding bit state either $|\varphi_1\rangle_{12}$ or $|\varphi_2\rangle_{12}$, respectively. If Eve detects inconclusive outcome then she is trying to guess possible Alice's signal and to masquerade as Alice. Consider it in detail on example of the state $\rho_1$. Assume that Eve resends a state $|\varphi_1\rangle_{12}$ with probability $P_1^{"}$ and $|\varphi_1^{'}\rangle_{12}$ with probability $P_2^{"}$ ($P_1^{"} + P_2^{"} = 1$) in the case of her inconclusive output. Then, Eve affects output probability distribution as $P_{0E}^{(Out)} = P_1\left(1 + P_1^{"}\right)/8 + P_1^{'}P_1^{"}/4$, where we even neglected $P_0\left(\alpha^{'}\right)$, where $\alpha^{'}$ is the amplitude of the displaced states that Eve creates. In general, Eve may choose $P_1^{"}$ in such a way that $P_{0E}^{(Out)}$ was almost similar to $P_0^{(Out)}$ (Eq. (6a)) due to the contribution $P_1^{'}P_1^{"}/4$ (she may sometimes guess correct distribution $P_0^{(Out)}$). But it happens at the expense of nonzero disguised probability $P_d = P_1^{'}P_1^{"}/4 \neq 0$ thus giving Eve's presence away. The more $P_1^{"}$ Eve chooses the more disguised probability $P_d$ is observed.

Eve may choose more tricky strategy of eavesdropping. Assume that Eve resends a corresponding disguised state either $|\varphi_1^{'}\rangle_{12}$ or $|\varphi_2^{'}\rangle_{12}$ if she has got corresponding inconclusive output but she resends the following states

$$|\Psi_1\rangle_{12} = \frac{1}{\sqrt{2}}\left(\left|1,\alpha^{'}\right\rangle_1\left|0,i\alpha^{'}\right\rangle_2 - i\left|0,\alpha^{'}\right\rangle_1\left|1,i\alpha^{'}\right\rangle_2\right), \tag{7a}$$

$$|\Psi_2\rangle_{12} = \frac{1}{\sqrt{2}}\left(-i\left|1,i\alpha_1^{'}\right\rangle_1\left|0,\alpha_1^{'}\right\rangle_2 + \left|0,i\alpha_1^{'}\right\rangle_1\left|1,\alpha_1^{'}\right\rangle_2\right), \tag{7b}$$

instead of $|\varphi_1\rangle_{12}$ or $|\varphi_2\rangle_{12}$, respectively, if she obtains a bit outcome. Such a strategy gives correct output distribution between Alice and Bob (6a)-(6c) since

$$\hat{B}_1|\Psi_1\rangle_{12} = |1\rangle_1\left|0,i\sqrt{2}\alpha^{'}\right\rangle_2 \tag{8a}$$

$$\hat{B}_1|\Psi_2\rangle_{12} = \left|0,i\sqrt{2}\alpha^{'}\right\rangle_1|1\rangle, \tag{8b}$$

saving for difference between $P_0(\alpha)$, $P_0(\alpha_1)$, $P_0(\alpha^{'})$ and $P_0(\alpha_1^{'})$. Then, Eve may share 1 bit of information with Alice and Bob. Nevertheless, such method of eavesdropping has a weak place. The states $|\Psi_1\rangle_{12}$ and $|\Psi_2\rangle_{12}$ are sensitive to influence of decoherence. It is impossible to keep the phase relation in the states $|\Psi_1\rangle_{12}$, $|\Psi_2\rangle_{12}$ stable when Eve and Bob are separated by large distance since quantum coherence is fragile under unavoidable interaction with environments. The decoherence effects for a state described by the density operator can be induced by solving the master equation when it is possible exactly calculate the coherence parameter and amplitude damping. Calculations of the parameters for the states (7a) and (7b) are beyond our consideration. Nevertheless, we may conjecture that Bob obtains a mixture of the states with density matrix

$\rho_1^{'} = 0.5\left(\left(|1,\alpha^{'}\rangle_{11}\langle 1,\alpha^{'}|\right)\otimes\left(|0,i\alpha^{'}\rangle_{22}\langle 0,i\alpha^{'}|\right)+\left(|0,\alpha^{'}\rangle_{11}\langle 0,\alpha^{'}|\right)\otimes\left(|1,i\alpha^{'}\rangle_{22}\langle 1,i\alpha^{'}|\right)\right)$ by analogy with coherent states with different amplitudes. Such a density matrix introduces error in output distribution $P_{0E}^{(Out)} = P\left(1-P_0\left(\alpha^{'}\right)\right)/8$, $P_{1E}^{(Out)} = P\left(1-P_0\left(\alpha_1^{'}\right)\right)/8$, and $P_{?E}^{(Out)} = 1-P_{0E}^{(Out)}-P_{1E}^{(Out)}$ on compared with Eqs. (6a)-(6c) that can be observed. It is possible to show that, when Eve eavesdrops on a fraction $\eta \le 1$ of the transmissions, then the final Alice-Bob distribution $P_{0E}^{(Out)} = P_{1E}^{(Out)} = P(1-\eta/2)(1-P_0(\alpha))/4$ and $P_{?E}^{(Out)} = 1-P_{0E}^{(Out)}-P_{1E}^{(Out)}$, provided that $P_0(\alpha) = P_0(\alpha^{'}) = P_0(\alpha_1^{'})$ is performed, may approach ideal one given by Eqs. (6a)-(6c) at the expense of 1 bit less of mutual information ($I(A,E) = I(E,B) = \eta$).

Let us consider another realistic strategy (beam splitting attack) where Eve tries to eavesdrop the transmitted signals without observing. Assume that Eve splits both states using her two beam splitters both described by the matrix

$B_E = \begin{bmatrix} T & R \\ -R^* & T^* \end{bmatrix}$, where $T$ and $R$ satisfy the condition $|T|^2+|R|^2=1$. Then, the output states are the following

$$\hat{U}\left(|1,\alpha\rangle_1|0,i\alpha\rangle_2\right)=\hat{B}_{1E_1}\left(|1,\alpha\rangle_1|0\rangle_{E_1}\right)\hat{B}_{1E_2}\left(|0,i\alpha\rangle_1|0\rangle_{E_2}\right)=T|1,\alpha T\rangle_1|0,i\alpha T\rangle_2|0,\alpha R\rangle_{E_1}|0,i\alpha R\rangle_{E_2}$$
$$+R|0,\alpha T\rangle_1|0,i\alpha T\rangle_2|1,\alpha R\rangle_{E_1}|0,i\alpha R\rangle_{E_2}, \quad (9a)$$

$$\hat{U}\left(|0,\alpha\rangle_1|0,i\alpha\rangle_2\right)=\hat{B}_{1E_1}\left(|0,\alpha\rangle_1|0\rangle_{E_1}\right)\hat{B}_{1E_2}\left(|0,i\alpha\rangle_1|0\rangle_{E_2}\right)=|0,\alpha T\rangle_1|0,i\alpha T\rangle_2|0,\alpha R\rangle_{E_1}|0,i\alpha R\rangle_{E_2}, \quad (9b)$$

where $E_1$ and $E_2$ are the Eve's modes. As $|R|<<1$, Eve may neglect contribution of second term in Eq. (9a) for her estimations. The same is applicable to the components of $\rho_2$. The best that Eve can do in the case is to choose the parameters of her beam splitters such the condition $|T|>>|R|$ to be performed. Then, the output Alice-Bob statistics $P_{0E}^{(Out)}=P|T|^2\left(1-P_0(\alpha T)\right)/4$, $P_{1E}^{(Out)}=P|T|^2\left(1-P_0(\alpha T)\right)/4$, $P_{?}^{(Out)}=1-P_{0E}^{(Out)}-P_{1E}^{(Out)}$ approaches sufficiently close to ideal (6a)-(6c), since $|T|^2\cong 1$. Alice and Bob compare their statistics and take it as correct, after that Bob announces the corresponding number where he has got bit value. Eve also listens to their talk and she needs only to distinguish two states $|0,\alpha R\rangle_{E_1}|0,i\alpha R\rangle_{E_2}$ and $|0,i\alpha_1 R\rangle_{E_1}|0,\alpha_1 R\rangle_{E_2}$ from each other to have an access to the coding. It can be done as it does Bob with help of the balanced beam splitter (3) $\hat{B}_1\left(|0,\alpha R\rangle_{E_1}|0,i\alpha R\rangle_{E_2}\right)=|0\rangle_1|0,i\sqrt{2}\alpha R\rangle_2$ and $\hat{B}_1\left(|0,i\alpha_1 R\rangle_{E_1}|0,\alpha_1 R\rangle_{E_2}\right)=|0,i\sqrt{2}\alpha_1 R\rangle_1|0\rangle_2$. Nevertheless, such strategy does not give Eve sufficient access to coding since a probability $P_{vac}=\exp\left(-2|\alpha|^2\left(1-|T|^2\right)\right)\approx 1$ not to register any photons and distinguish between $|0,\alpha R\rangle_{E_1}|0,i\alpha R\rangle_{E_2}$ and $|0,i\alpha_1 R\rangle_{E_1}|0,\alpha_1 R\rangle_{E_2}$, respectively, is high. Eve registers nothing and she loses any information about coding shared by Alice and Bob. Thus, she may only has access to $\eta=1-P_{vac}\approx 0$ bits of mutual information. Moreover, Eve does not know exactly values of $\alpha$ to try to define optimal parameters for her beam splitting attack. This consideration gives estimations for Alice's amplitude $\alpha$ to satisfy condition $|\alpha|^2\left(1-|T|^2\right)\cong 0$ for $|T|^2\cong 1$.

Now, consider the case when Eve attempts to gain some information on each signal sent by Alice, while minimizing the damage to the state. This strategy can be realized by making the information carrier interact unitarily with a probe, and then letting it proceed to Bob, in slightly modified state. Eve may store her probe and decides which type of measurement to perform on her probe only after Alice and Bob shared their coding. To do it Eve supplies her probe in a known initial state $|g\rangle$, and then the combined system may evolve as

$$\hat{U}\left(|\varphi_1\rangle|g\rangle\right)=|\varphi_{1E}\rangle|e_1\rangle, \qquad \hat{U}\left(|\varphi_1^{'}\rangle|g\rangle\right)=|\varphi_{1E}^{'}\rangle|e_2\rangle, \tag{10a}$$

$$\hat{U}\left(|\varphi_2\rangle|g\rangle\right)=|\varphi_{2E}\rangle|e_3\rangle, \qquad \hat{U}\left(|\varphi_2\rangle|g\rangle\right)=|\varphi_{2E}^{'}\rangle|e_2\rangle, \tag{10b}$$

where $|\varphi_{1E}\rangle_{12}=|1,\alpha_E\rangle_1|0,i\alpha_E\rangle_2$, $|\varphi_{1E}^{'}\rangle_{12}=|0,\alpha_E\rangle_1|0,i\alpha_E\rangle_2$, $|\varphi_{2E}\rangle_{12}=|1,i\alpha_{1E}\rangle_1|0,\alpha_{1E}\rangle_2$, and $|\varphi_{2E}^{'}\rangle_{12}=|0,i\alpha_{1E}\rangle_1|0,\alpha_{1E}\rangle_2$. Evolution is unitary (Eve can to make some Hamiltonian which generates it) and scalar product is conserved. Then, it imposes the following condition

$$\langle e_1|e_3\rangle=\exp\left(-\left(|\alpha|^2-|\alpha_E|^2\right)-\left(|\alpha|^2-|\alpha_{1E}|^2\right)\right)\frac{1-(i\alpha-\alpha)^2}{1-(i\alpha_{1E}-\alpha_E)^2}, \tag{11a}$$

$$\langle e_1|e_4\rangle=\exp\left(-\left(|\alpha|^2-|\alpha_E|^2\right)-\left(|\alpha|^2-|\alpha_{1E}|^2\right)\right)\frac{i\alpha-\alpha}{i\alpha_{1E}-\alpha_E}, \tag{11b}$$

$$\langle e_3|e_2\rangle=\exp\left(-\left(|\alpha|^2-|\alpha_E|^2\right)-\left(|\alpha|^2-|\alpha_{1E}|^2\right)\right)\frac{\alpha-i\alpha}{\alpha_E-i\alpha_{1E}}, \tag{11c}$$

$$\langle e_4|e_2\rangle=\exp\left(-\left(|\alpha|^2-|\alpha_E|^2\right)-\left(|\alpha|^2-|\alpha_{1E}|^2\right)\right). \tag{11d}$$

The composite system is a direct product of the corresponding states if overlaps $\left|\langle e_i|e_j\rangle\right|^2\le 1$ ( $i,j=1-4$ ). After sending the modified carrier to Bob, Eve remains with her probe. The probes are not orthogonal to each other. The idea of Eve is to cause minimal damage to the

information carrier and to obtain as much as possible information. To hide her presence Eve may try to guess Alice's parameters $\alpha \cong \alpha_E$ and $\alpha \cong \alpha_{1E}$ to provide performance of condition $P_0(\alpha) \approx P_0(\alpha_E) \approx P_0(\alpha_{1E})$. But overlapping $\langle e_1 | e_3 \rangle$ (Eq. (11a)) becomes almost unit ( $\langle e_1 | e_3 \rangle \approx 1$ ) for the case of $\alpha \cong \alpha_E$ and $\alpha \cong \alpha_{1E}$, respectively. Since the states $|e_1\rangle$ and $|e_3\rangle$ are not orthogonal and even more their overlapping is sufficiently large, Eve cannot distinguish them exactly and, as consequence, she may share only 1 bit less of mutual information.

**4. Discussion and conclusion**

Optical quantum cryptography is based on the use of single photon Fock states. Unfortunately, these states are difficult to realize experimentally. Nowadays, practical implementations rely on faint laser pulses, in which the photon number distribution obeys Poisson statistics or entangled photon pair. Both the possibilities suffer from a small probability of generating more than one photon or photon pair at the same time. For large losses in the quantum channel, small fractions of these multiphotons can have important consequences on the security of the key. We propose not to pursue goal of creating ideal resource of single photon states and make use of really existing resource of single photons. The way to create pseudo-single-photon states is the generation of photon pairs and the of use one photon as a trigger for the other one. The conditional photon generation by parametric down converter is connected with imperfections associated with experimental technique results [17]. Nevertheless, if we modulate such a statistical mixture by coherent state on a beam splitter we produce displaced photon states that are applicable for proposed QKD protocol. Even more, developed QKD protocol based on use of dual rail displaced states works with experimental attendant noise used to observe possible presence of malicious Eve.

By the way, such a modulation enables Alice to use two additional parameters inaccessible to nobody, namely, her initial distributions between displaced single photon and vacuum and amplitudes of her fields which she may change. Such QKD deals with optical pulses as carriers unlike quantum QKD with a single photon that approximates it to standard telecommunication communication. With the availability of source of quantum states for the communication, the success of quantum cryptography essentially depends on the ability to detect single photons. In principle, this can be achieved using a variety of techniques, for instance, photomultipliers, avalanche photodiodes, multichannel plates, and superconducting Josephson junctions. In our case, we need only to distinguish between a single photon and optical pulses involving multiphoton states that can be done by means of commercial detectors as it is shown in Fig. 1 and 2.

Note another peculiarity of the proposed scheme. Consider optical fiber version of a Mach-Zehnder interferometer made out of two symmetric beam splitters connected to each other, with one phase modulator in each arm. This interferometer combined with a single-photon source and photon-counting detectors can be used for quantum cryptography provided that phase shift is kept constant. Although such a scheme may be perfect on an optical table, it is impossible to keep the path difference between two modes stable for a distance more. If we take states similar to 7(a) and 7(b) as carriers, then the same problems appear as the states 7(a) and 7(b) are the displaced analogues of a singe photon in superposition state that takes simultaneously two modes. Nevertheless, although we call used states as dual rail displaced photon number states, it is evident that Alice can do some delay between the pulses in different modes and send them through the same optical fiber one after the other where they may experience the same phase shift in environmentally sensitive part of the system. This enables to conserve phase relations of incoming pulses on output at Bob side if he also makes the same delay for first pulse before to combine two pulses on the beam splitter. Detailed analysis of influence of decoherence on phase relations is subject of future investigation and is beyond the consideration. Let us only mention that use of pulses with large amplitudes unlike conventional schemes of the quantum cryptography may show resistance to eavesdropping even in settings with high attenuation. It is also useful to note that optical scheme of two-state protocol [5] can be implemented using interference between a macroscopic bright pulse and a dim pulse with less than one photon on average [5]. Proposed optical scheme is not one of Mach-Zehnder interferometer and, as consequence, it is free of interference effect and of attendant problems. Remarkably, that this approach is robust against loss of the single photon, and inefficiency of the photodetectors. Those factors will cause the corresponding photodetectors to be silent, and such cases can simply be discarded. Therefore, this only affects the output distributions and has to be taken into account in real case.

In conclusion, we proposed a new QKD protocol that is based on use of nonclassical properties of the displaced single photon states. Given protocol works as a binary erasure channel also as in a B-92 protocol [5]. This is sole possible resemblance with a B-92 protocol but not more. Our analysis involves study of only restricted number of possible eavesdropping attacks and show that the protocol is secure under them.

**Acknowledgment**

This work was partially supported by the IT R&D program of MKE/IITA (2008-F-035-01).

**List of figures**

**Figure 1**

Schematic representation of QKD based on dual-rail displaced states. Alice prepares her dual-rail displaced state and sends it to Bob who has a chance extract bit value if it was the bit state. Otherwise Bob obtains inconclusive outcome and discards it. Bob announces a number where he successfully got bit value only if a procedure of check of eavesdropping showed absence of it.

**Figure 2(a,b)**

Example how to distinguish between a coherent state $\left|0,\sqrt{2}\alpha\right\rangle$ and a single photon $\left|1\right\rangle$. The coherent state mainly results in registration of photons by two detectors save for small failure probability to register only one click. Single photon gives always one click. The more amplitude of the displaced state we use the less the failure probability.

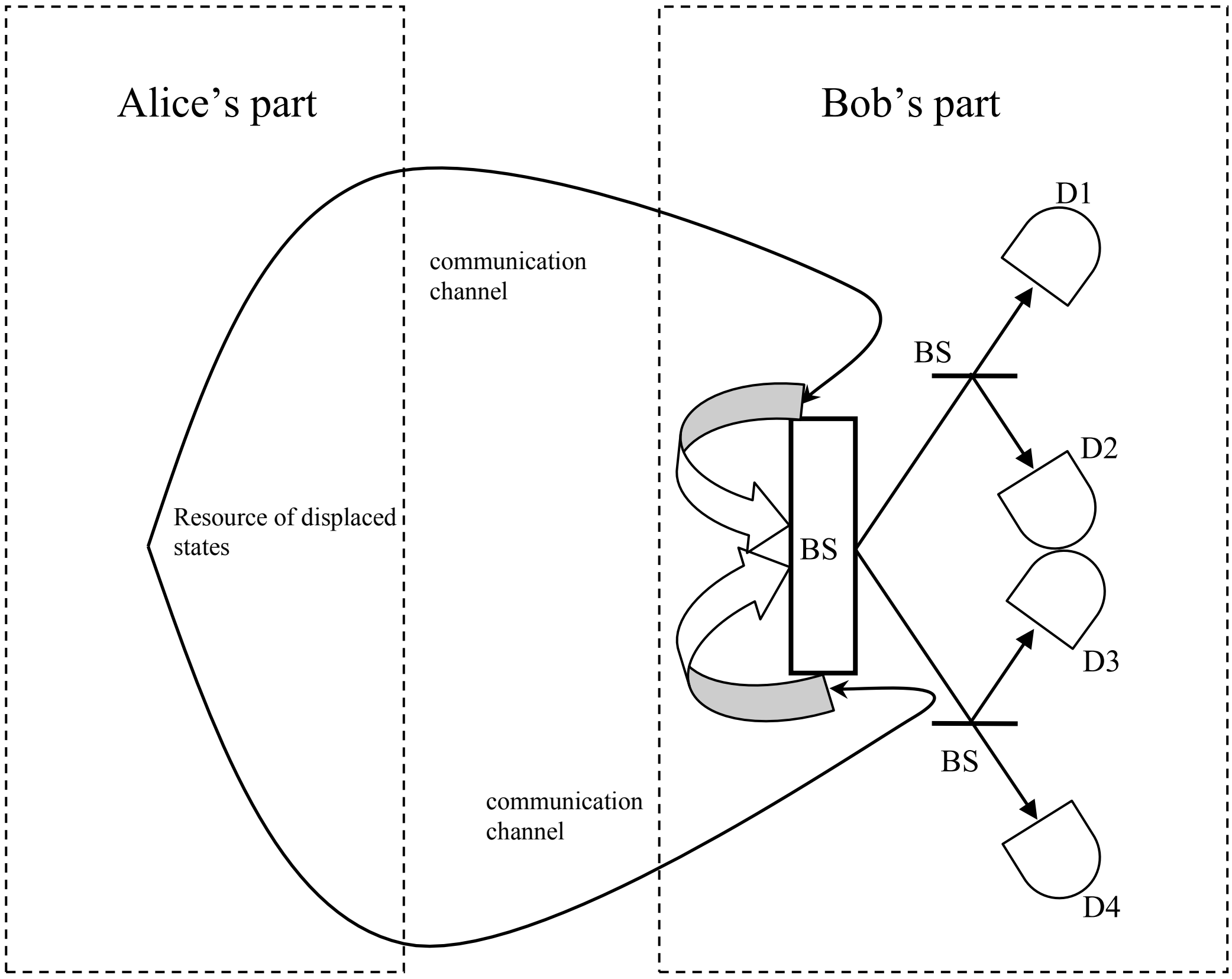
Alice's part
Bob's part
communication channel
Resource of displaced states
BS
BS
BS
D1
D2
D3
D4
communication channel

**Figure 1**

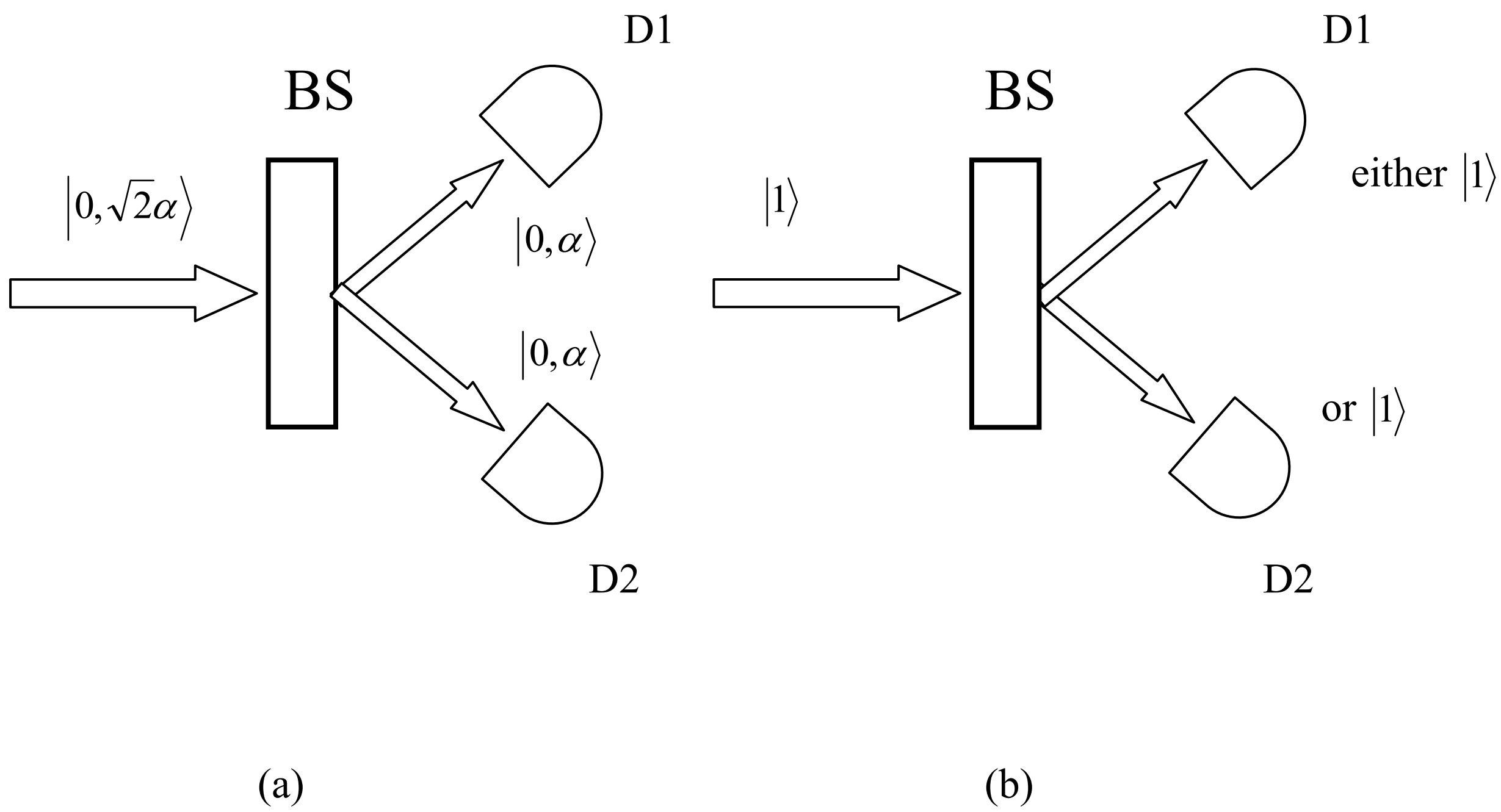
D1
BS
$|0,\sqrt{2}\alpha\rangle$
$|0,\alpha\rangle$
$|0,\alpha\rangle$
D2
D1
BS
$|1\rangle$
either $|1\rangle$
or $|1\rangle$
D2
(a)
(b)

**Figure 2 (a, b)**